\documentclass{article}
\usepackage{graphicx}
\usepackage[hang]{caption}
\usepackage{subcaption} 
\usepackage[utf8]{inputenc}
\usepackage[T1]{fontenc}
\usepackage[english]{babel}
\usepackage{microtype} % optional, for aesthetics
\usepackage{tabularx} % nice to have
\usepackage{booktabs} % necessary for style
\usepackage{graphicx}
\usepackage{indentfirst}
\usepackage{amssymb}
\usepackage{amsmath}

\usepackage{bm}
\usepackage{braket}
\usepackage[margin=1cm]{caption}

\title{Spacetime Path Integrals for Entangled States}
\author{Narayani Tyagi, Ken Wharton \\ 
Department of Physics and Astronomy \\
San Jos\'{e} State University, San Jos\'{e}, CA 95192-0106}
\date{}
\begin{document}
\maketitle

\begin{abstract}
Although the path-integral formalism is known to be equivalent to conventional quantum mechanics, it is not generally obvious how to implement path-based calculations for multi-qubit entangled states.  Whether one takes the formal view of entangled states as entities in a high-dimensional Hilbert space, or the intuitive view of these states as a connection between distant spatial configurations, it may not even be obvious that a path-based calculation can be achieved using only paths in ordinary space and time.  Previous work has shown how to do this for certain special states; this paper extends those results to all pure two-qubit states, where each qubit can be measured in an arbitrary basis.  Certain three-qubit states are also developed, and path integrals again reproduce the usual correlations.  These results should allow for a substantial amount of conventional quantum analysis to be translated over into a path-integral perspective, simplifying certain calculations, and more generally informing research in quantum foundations.
\end{abstract}

\section{Introduction} \label{Introduction}

Entangled $N$-particle quantum states live in a high-dimensional Hilbert space or configuration space, with no evident representation on ordinary space or spacetime.  Even when limiting analysis to which-way entanglement -- a case seemingly aligned with a spatial representation -- quantum theory formally uses a superposition of spatial configurations, with the total wavefunction therefore defined as a function on a $3N$-dimensional configuration space $\psi(\bm{x}_1,\bm{x}_2,\dots,\bm{x}_N)$ rather than a set of $N$ single-particle functions on ordinary 3D space, $\psi_n(\bm{x})$.  In fact, the hallmark of entangled states is that they \textit{cannot} be represented as functions on ordinary spacetime.  This disconnect between wavefunctions and conventional spacetime not only makes it difficult to visualize quantum states, but is also arguably a fundamental source of conflict between quantum theory and general relativity. \cite{Kent2014}

There is a mistaken but commonly-held view that the disconnect between quantum theory and spacetime also applies to the Feynman path integral \cite{Feynman1948}.  In this view, analyzing entangled systems via the path integral requires one to consider all possible ``paths'' through Hilbert space, including ``paths'' that consist of pure entangled states.  Because a single entangled state has no spatial representation, any of these ``paths'' would also have no possible spacetime representation.  Approaches using such Hilbert space ``paths'' are common in the literature \cite{Cotler2016,Farhi1992}, and have even been extended to quantum field theory \cite{Green2016}.  

The central result of this paper is to demonstrate how the path integral can be applied to general entangled states without requiring a departure from a conventional spatial representation.  In other words, each relevant step in the calculation can correspond to \textit{physical} paths through ordinary spacetime, with no need for scare-quotes around ``path''.  The success of this procedure is most evident for the case of which-way entanglement.  One need simply use the basis of the spatial configuration space $\psi(\bm{x}_1,\bm{x}_2,\dots,\bm{x}_N)$, and perform the path integral in this basis.  Any point on one of these paths is simply a configuration of spatial particle positions, representing each particle at a position in ordinary 3D space (the same way configuration space is used in classical statistical mechanics).  Any single path through points in 3N-dimensional configuration space, therefore, corresponds to N paths through conventional spacetime.  We will follow standard terminology in calling this collection of $N$ paths a single ``history'' \cite{sinha1991}.  This preference for the position basis was advocated by Feynman's original paper \cite{Feynman1948}, and has been utilized in other attempts to have a realistic interpretation of the path integral \cite{Kent2013,Sorkin2007,Wharton2016}.

A few explicit examples of spacetime-based path integrals for entangled states have already appeared in the literature \cite{sinha1991,Wharton2011}, but these only cover very special states (maximally entangled 2-qubit Bell states).  No paper has yet addressed how to perform this analysis for partially-entangled states or 3-qubit GHZ states; these cases are developed in detail below.  As in the previous work, we indicate how to simplify this analysis by only needing to consider a few classical paths, rather than also summing over all non-classical paths.  (In many known cases, all non-classical paths end up either cancelling exactly or simply reinforcing classical paths \cite{Schulman2005}; fortunately the same simplification works correctly in the cases considered here.)

The techniques developed below are evidently generalizable to more complicated entangled states.  Quantum information theory has a substantial number of tools which allow one to convert any conceivable entangled state into a physical experimental geometry -- translating any sort of entanglement into which-way entanglement -- for which this general approach could be utilized.\footnote{There are reasonable arguments as to why the approach here could also be extended to other types of entanglement, but for the purposes of this paper, we shall focus on path integral accounts of which-way entanglement.} 

There are a number of possible motivations for taking an abstract high-dimensional entangled state and analyzing it via paths in conventional 4D spacetime.  For many physicists, the main advantage might simply be a matter of ease of visualization.  A generic three-qubit state lives in an 8-dimensional complex space, or, if normalized, as a point on a real 15-dimensional sphere.  Analyzing such states via paths through ordinary space is evidently a preferable representation.\footnote{Granted, many physicists commonly think of entangled states as ``superpositions'' of configurations in ordinary space, which is arguably comparable to a set of possible paths.  But apart from the disconnect between this viewpoint and the formal mathematics, this model of entangled states further requires the visualization of a mysterious direct connection between spacelike-separated events, a connection not required in the path integral.} 

Another set of motivations will be evident to those in the field of quantum foundations.  Many key theorems and discussions involve entangled states, but it is usually not clear how to translate those results into a path-integral framework, missing out on a potentially useful additional perspective.  Evidently, such a translation would also be critical for research programs which aim to develop a realistic account of path integrals  \cite{Kent2013,Sorkin2007,Wharton2016}.  The mere fact that the path integral mathematics is framed on a spacetime background makes it more suitable for any analysis of entangled states in curved spacetime.  And since the path integral solves systems ``all at once'' rather than via dynamical evolution, it ties directly into the main remaining approach for finding spacetime-based accounts of quantum phenomena \cite{Wharton2014,Wharton2020}.  Any such analysis would necessarily require some way to translate between entangled states and physical paths through spacetime.  The results in this paper indicate how such a translation might be accomplished.

The next section focuses on a which-way representation of a single qubit and develops a single qubit measurement device. This device is a key part of the experimental geometries used in rest of the paper.  Section \ref{The 2 Qubit case} delves into the observable correlations of the entangled two-qubit state. A generic two qubit state is mapped to an experimental setup that is amenable to a path-integral analysis.  The basic rules for the path integral calculations are given in Section \ref{rules for PI}, which are then applied to this experiment.  The results exactly compare to the usual quantum probabilities.  Then, in Section \ref{The 3 Qubit case: GHZ}, we develop the specific case of the three-qubit GHZ state \cite{greenberger1990}. Again, the path integral calculation is successful.  Finally, these results are discussed in Section \ref{Discussion and Conclusion}, including future applications. 

%\renewcommand*\contentsname{Outline}
%\tableofcontents

\section{Single Qubit Measurements} \label{Single Qubit Measurements}

While there are many experimental realizations of any given entangled state, the approach in this paper is to utilize photons rather than massive particles, interacting exclusively with beamsplitters, phase-shifting devices (phase plates), and single-photon detectors.  (The sources of these photons will be developed in section 3.) 

In this section we focus on a which-way representation of a single qubit, and demonstrate how to build a device which can serve as a generic single qubit measurement.  Specifically, we would like a device that could be adjusted to measure a qubit on any chosen basis (aligned in any chosen direction on the Bloch sphere).  As we shall see, once the measurement basis is imposed via beamsplitters and phase plates, paths through this measurement device will enter into the path integral calculations in the next section.  

 \subsection{Quantum Probabilities} \label{Quantum Probabilities}
 
The most straightforward path-based implementation of a generic qubit $\ket{\psi}=\bm{a}\ket{0}+\bm{b}\ket{1}$ is to have $\ket{0}$ correspond to one path and $\ket{1}$ correspond to another.  The goal is to design a measurement device that can effectively project this qubit onto an arbitrary basis using beamsplitters and phase plates.  Using the usual spherical coordinates, this chosen basis is defined on the Bloch sphere via adjustable variables $\phi$ and $\theta$.  Specifically, this detector should project the qubit into the chosen basis $\ket{\large{\bm{e_+}}}$  and $\ket{\large{\bm{e_-}}}$, where
    \begin{equation}
       \ket{\large{\bm{e_+}}} = 
       \begin{pmatrix} 
         cos \frac{\theta}{2}\\
         sin \frac{\theta}{2}\: e^{i\phi}
        \end{pmatrix},
        \label{eqn:mmt at +}
    \end{equation} 
      
    \begin{equation}
      \ket{\large{\bm{e_-}}} =  \begin{pmatrix} 
         sin \frac{\theta}{2}\: e^{-i\phi}\\-\: cos \frac{\theta}{2}
       \end{pmatrix}.
     \end{equation}
    
  This is written in the usual spinor notation, where $\ket{0}$ is the top term and $\ket{1}$ is the bottom.  From the Born rule, the probability for a $\bm{+}$ outcome (corresponding to $\ket{\large{\bm{e_+}}}$) should be
    \begin{eqnarray}
    \label{eq:probplus}
      \left|\braket{\large{\bm{e_+}}|\large\psi}\right|^2
        %= |\: (\begin{pmatrix} 
          %   cos \frac{\theta}{2} & sin \frac{\theta}{2}\: e^{-i\phi}
           % \end{pmatrix})(\begin{pmatrix} a\\b\end{pmatrix})\: |^2 %\nonumber \\
      = \left|\: \bm{a}\: cos\frac{\theta}{2} + \bm{b}\: sin\frac{\theta}{2}\: e^{-i\phi}\: \right|^2.
    \end{eqnarray} 
    Similarly, the probability for the $\bm{-}$ outcome should be
     \begin{eqnarray}
     \label{eq:probminus}
      \left|\braket{\large{\bm{e_-}}|\large\psi}\right|^2 = \left|\: \bm{a}\: sin\frac{\theta}{2}\: e^{i\phi}\: - \: \bm{b}\: cos\frac{\theta}{2}\:  \right|^2.
     \end{eqnarray} 
     
\subsection{Measurement Device} \label{Measurement Device}     

To design a device which will yield the above results, it is useful to apply the correspondence principle that relates classical electromagnetic (EM) intensities with photon detection probabilities.  With this in mind, consider replacing the single photon with two classical EM plane waves, one on path $\ket{0}$ and one on path $\ket{1}$.  The electric field of each of these EM waves has an amplitude and a phase, and they can be expressed as complex field amplitudes $E_a$ and $E_b$.  Normalizing the total intensity $|{E_a}|^2 + |{E_b}|^2 = 1$ allows one to relate these waves to the original qubit $\ket{\psi}=\bm{a}\ket{0}+\bm{b}\ket{1}$.  In other words, if one wave is assigned a complex amplitude of $E_a=\bm{a}$, and the other one $E_b=\bm{b}$, they encode the same information as the original qubit.  (These classical EM waves are assumed to have the same frequency and polarization, to allow for perfect interference.)  

\begin{figure}[ht]
    \centering
      \includegraphics[width=0.5\textwidth,height=5cm]{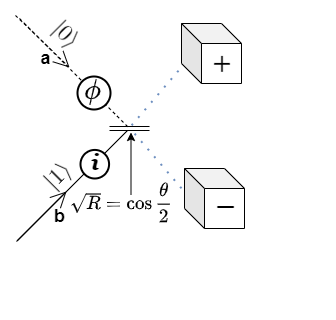}

    \caption{A generic measurement device for a qubit, where the superposition of the two input paths corresponds to a single photon.  An adjustable beamsplitter and phase plates project the qubit onto an arbitrary direction on the Bloch Sphere.} 
    \label{fig:1q}
  \end{figure}

  Now consider the measurement device shown in Figure \ref{fig:1q} with two inputs corresponding to paths $\ket{0}$ and path $\ket{1}$. As explained above, these inputs can correspond to classical EM waves with complex field amplitudes $\bm{a}$ and $\bm{b}$.  We can now calculate the intensities at the two detectors ( $\bm{+}$ and $\bm{-}$), corresponding to the two possible outcomes in the chosen measurement basis).  If we design a device where the intensities properly correspond to the quantum probabilities, it follows that this device is guaranteed to perform the desired measurement for the corresponding single photon.

     The desired measurement device consists of an adjustable phase plate adding a phase angle $\phi$ to the input $\ket{0}$, and a fixed phase plate adding a phase angle $\pi/2$ to the input $\ket{1}$.  The latter phase plate is indicated by ``$i$'', since this is the net complex phase factor $exp(i{\pi}/{2})=i$.  After the phase shifts, these two waves encounter a lossless beamsplitter, with a reflectivity factor corresponding to the adjustable variable $\theta$.  The reflected intensity factor is $R$, but the reflected field amplitude is scaled by $\sqrt{R}$, where
     \begin{eqnarray}
         \sqrt{R}=\cos\frac{\theta}{2} ,\\
         \sqrt{T}=\sin\frac{\theta}{2} .
    \end{eqnarray}
    such that $R + T = 1$.

   First we will calculate the \textit{total} intensity at the $\bm{+}$ detector.  As the input wave along path $\ket{0}$ passes through the adjustable phase plate $\bm{\phi}$, the amplitude $\bm{a}$ is multiplied by the phase factor $exp(i\phi)$ . At the beamsplitter, this wave gets both transmitted and reflected. However, we're interested in the reflected part that reaches the $\bm{+}$ detector. At the beamsplitter, the reflected component picks up a $\pi/2$ phase shift relative to the transmission \cite{degiorgio1980, zeilinger1981}, corresponding to a net factor $i\sqrt{R}$ due to the reflection. Thus this component of the wave arriving at the $\bm{+}$ detector has an amplitude of $i \bm{a}\sqrt{R}exp(i\phi)$.

    For the input wave travelling along $\ket{1}$, the amplitude $\bm{b}$ is multiplied by a phase factor $i$ due to the phase plate, and a transmission factor \textit{$\sqrt{T}$} from the beamsplitter (to reach the $\bm{+}$ detector) yielding an amplitude of $i\bm{b}\sqrt{T}$ at the detector.
 
      Thus, the total intensity at the $\bm{+}$ detector is the square of the sum of field amplitudes travelling along the $\ket{0}$ and $\ket{1}$ paths and can be given as:
   
    \begin{eqnarray}
         I_+=\left|i\bm{a}\: cos\frac{\theta}{2}\: e^{i\phi} + i\bm{b}\: sin\frac{\theta}{2}\right|^2 .
         \label{eqn:intensity+}
    \end{eqnarray}

    Recall that the sum of intensities has been normalized to be equal to 1. With the normalized intensity, this result is equivalent to Eqn. (\ref{eq:probplus}).
    
    A similar calculation can be done to find the intensity at the $\bm{-}$ detector.  The input along path $\ket{0}$ passes through the phase plate $\phi$ and reaches the beamsplitter.  In this case we are interested only in the transmitted wave that reaches the $\bm{-}$ detector, for another factor of $\sqrt{T}$.  The input wave along path $\ket{1}$ passes through the phase plate $i$, but only the reflected wave at the beamsplitter reaches the $\bm{-}$ detector, for a factor of $i\sqrt{R}$. Together, the intensity at the $\bm{-}$ detector is therefore:
    \begin{eqnarray}
          I_-=\left|\bm{a}\: sin\frac{\theta}{2}\: e^{i\phi}\: + \: i^2\: \bm{b}\: cos\frac{\theta}{2}\right|^2 .
        %   \left|\: \bm{a}\: sin\frac{\theta}{2}\: e^{i\phi}\: - \: \bm{b}\: cos\frac{\theta}{2}\:  \right|^2
        \label{eqn:intensity-} 
    \end{eqnarray}
    Once again, as the sum of intensities has been normalized to 1, this result is equivalent to Eqn. (\ref{eq:probminus}).

From the above results, we can see that the device pictured in Figure 1 is equivalent to a measurement of a qubit $\ket{\psi}=\bm{a}\ket{0}+\bm{b}\ket{1}$, where the amplitude of the upper-left input in Figure 1 corresponds to the $\ket{0}$ state and the amplitude of the lower-left input corresponds to $\ket{1}$. In the single photon limit, this is equivalent to measuring the qubit in a basis chosen by the angles $(\theta,\phi)$.
   
\section{The 2-Qubit Path Integral}\label{The 2 Qubit case}
  
   \subsection{Quantum Mechanical probabilities}\label{Quantum Mechanical probabilities}

   The main result of this paper is to show that the observable correlations of any entangled two-qubit state can be computed from a path-integral perspective, using paths in physical spacetime.  Thanks to the Schmidt decomposition \cite{schmidt1907,peres2006,wharton2016b} -- see Appendix A for explicit details -- a generic two-qubit (pure) state can always be written in the simple form
  \begin{equation}
      \label{eq:psi2}
      \ket{\psi}=A\ket{0}\otimes\ket{1}+B e^{i\delta}\ket{1}\otimes\ket{0}=\begin{pmatrix} 0\\A\\B\:e^{i\delta}\\0 \end{pmatrix}.
  \end{equation}
  Here the constants $A$ and $B$ are assumed to be real and positive.  
  
  The basis for each qubit in the above expression is fixed to the Schmidt basis (defined in Appendix A), but the below results are still generally applicable to any two-qubit state written in any basis, from which the Schmidt basis can be calculated.  The general case would merely require particular rotations of the measurement coordinate systems, as precisely specified in Appendix A.  The main paper will now proceed assuming the translation of a generic state into the form in Eqn. (\ref{eq:psi2}) has already been accomplished, such that the measurement basis for each qubit is defined relative to the Schmidt basis used in (\ref{eq:psi2}).
  
  Using similar notation to the previous section, the two independent measurement bases are $\ket{\large{\bm{e_+}}}_1$ and $\ket{\large{\bm{e_-}}}_1$ for the first qubit, and $\ket{\large{\bm{e_+}}}_2$ and $\ket{\large{\bm{e_-}}}_2$ for the second qubit. On the Bloch sphere, $\ket{\large{\bm{e_+}}}_1$ corresponds to the freely chosen angles ($\theta_1,\phi_1$), and $\ket{\large{\bm{e_+}}}_2$ corresponds to ($\theta_2,\phi_2$).  The tensor product of these single-qubit bases leads to the following definitions, corresponding to the four possible measurement outcomes: 
   \begin{eqnarray}
    \ket{\bm{++}}=\ket{\large{\bm{e_+}}}_1\otimes \ket{\large{\bm{e_+}}}_2  = \begin{pmatrix} cos \frac{\theta_1}{2}\:cos\frac{\theta_2}{2}\\
    cos \frac{\theta_1}{2}\:sin\frac{\theta_2}{2}\:e^{i\phi_2}\\
    sin \frac{\theta_1}{2}\:cos\frac{\theta_2}{2}\:e^{i\phi_1}\\
    sin \frac{\theta_1}{2}\:sin\frac{\theta_2}{2}\:e^{i(\phi_1 + \phi_2)}\end{pmatrix} ,
   \end{eqnarray}
   
   \begin{eqnarray}
    \ket{\bm{+\:-}}=\ket{\large{\bm{e_+}}}_1\otimes{\ket{\large{\bm{e_-}}}}_2  = \begin{pmatrix} cos \frac{\theta_1}{2}\:sin\frac{\theta_2}{2}\:e^{-i\phi_2}\\-cos \frac{\theta_1}{2}\:cos\frac{\theta_2}{2}\\sin \frac{\theta_1}{2}\:sin\frac{\theta_2}{2}\:e^{i(\phi_1 - \phi_2)}\\-sin \frac{\theta_1}{2}\:cos\frac{\theta_2}{2}\:e^{i\phi_1}\end{pmatrix} ,
   \end{eqnarray}
  
  \begin{eqnarray}
    \ket{\bm{-\:+}}=\ket{\large{\bm{e_-}}}_1\otimes{\ket{\large{\bm{e_+}}}}_2  = \begin{pmatrix} sin \frac{\theta_1}{2}\:cos\frac{\theta_2}{2}\:e^{-i\phi_1}\\sin \frac{\theta_1}{2}\:sin\frac{\theta_2}{2}\:e^{i(\phi_2 - \phi_1)}\\-cos \frac{\theta_1}{2}\:cos\frac{\theta_2}{2}\\-cos \frac{\theta_1}{2}\:sin\frac{\theta_2}{2}\:e^{i\phi_2}\end{pmatrix} ,
  \end{eqnarray}
   
  \begin{eqnarray}
    \ket{\bm{-\:-}}=\ket{\large{\bm{e_-}}}_1\otimes{\ket{\large{\bm{e_-}}}}_2  = \begin{pmatrix} sin \frac{\theta_1}{2}\:sin\frac{\theta_2}{2}\:e^{-i(\phi_2 + \phi_1)}\\-sin \frac{\theta_1}{2}\:cos\frac{\theta_2}{2}\:e^{i\phi_1}\\-cos \frac{\theta_1}{2}\:sin\frac{\theta_2}{2}\:e^{-i\phi_2}\\cos \frac{\theta_1}{2}\:cos\frac{\theta_2}{2}\end{pmatrix}    .
  \end{eqnarray}
      Suppose we have an entangled quantum state of the form $\ket{\psi}$ as shown in Eqn. (\ref{eq:psi2}). Then, for the arbitrary measurement basis defined above, we can find the corresponding QM probabilities for the 4 outcomes as follows: %$\ket{+\:+}$, $\ket{+\:-}$, $\ket{-\:+}$ and $\ket{-\:-}$  
   \begin{eqnarray}
         \left|\braket{{+\:+}|\psi}\right|^2 &=& \left|A\:cos \frac{\theta_1}{2}\:sin \frac{\theta_2}{2}\:e^{-i\phi_2} + B\:e^{i\delta}sin \frac{\theta_1}{2}\:cos \frac{\theta_2}{2}\:e^{-i\phi_1}\right|^2 ,\label{eqn:qm++}\\
         \left|\braket{{+\:-}|\psi}\right|^2 &=& \left|-A\:cos \frac{\theta_1}{2}\:cos \frac{\theta_2}{2} + B\:e^{i\delta}sin \frac{\theta_1}{2}\:sin \frac{\theta_2}{2}\:e^{-i(\phi_1-\phi_2)}\right|^2 ,\label{eqn:qm+-}\\
         \left|\braket{{-\:+}|\psi}\right|^2 &=& \left|A\:sin \frac{\theta_1}{2}\:sin \frac{\theta_2}{2}\:e^{-i(\phi_2-\phi_1)} - B\:e^{i\delta}cos \frac{\theta_1}{2}\:cos \frac{\theta_2}{2}\right|^2 ,\label{eqn:qm-+}\\
         \left|\braket{{-\:-}|\psi}\right|^2 &=& \left|-A\:sin \frac{\theta_1}{2}\:cos \frac{\theta_2}{2}\:e^{-i\phi_1} - B\:e^{i\delta}cos \frac{\theta_1}{2}\:sin \frac{\theta_2}{2}\:e^{i\phi_2}\right|^2 .\label{eqn:qm--}
   \end{eqnarray}
   
   The goal of this section is to show that these results can be easily calculated in a spacetime-path integral framework.

      \begin{figure}[!ht]
        \centering
        \includegraphics[width=0.6\textwidth,height=10cm]{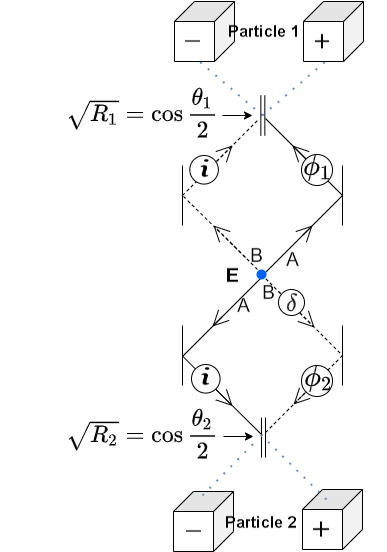}
        \caption{ The setup consists of two generic measurement devices(refer Figure \ref{fig:1q}), one for each qubit. Source \textbf{E} emits two photons travelling in opposite directions along solid or dashed paths. The first photon reaches the top set of detectors while the second photon reaches the bottom set of detectors. An adjustable beamsplitter and phase plates are associated with each measurement device, corresponding to a measurement basis that can be chosen independently for each qubit. }
        \label{fig:2q}
      \end{figure}
   
   \subsection{Two-Photon Experimental Geometry}\label{Two-Photon Experimental Geometry}

       From the results in section 2, we know that a which-way implementation of a generic two-qubit measurement will require a two-photon source, with each photon encountering one of the devices pictured in Figure 1.  Such an experiment is indicated in Figure \ref{fig:2q}.  Here we have a two-photon source $\bm{E}$ that emits two photons in opposite directions.\footnote{While there are various physical mechanisms that emit photons in opposite directions, further details of this source are not important for any of the calculations below, except we will need certain directions to be more probable than others.}  In the traditional quantum viewpoint, this implies a superposition of photon emissions on the solid and dashed lines of Figure \ref{fig:2q}, but such a viewpoint is not needed for the path integral approach.  Here, we take the photons to be constrained to follow \textit{either} the two solid lines \textit{or} the two dashed lines.  Another difference is that no collapse is needed; in any given history, each photon will either transmit or reflect from a beamsplitter, and will end up at only one particular detector.
      
     In order to have an analog to the most general two-qubit state where $A\ne B$, we require a source that preferentially emits in certain directions.  Imagine that the source was surrounded by a screen.  In this case, we require that the probability of emissions along the solid lines is $A^2$, and the probability along the dashed lines is $B^2$.  We ignore all other emission directions\footnote{If the photons are emitted in any other direction, they will not be detected, and will not enter into to the measurement probabilities.} such that $A^2+B^2=1$. Notice that one of the photons emitted along the dashed lines would pass through phase plate with phase angle $\delta$.  We can imagine that the set-up of the source $\bm{E}$ and the angle $\delta$ are under the control of an experimenter with the goal of providing two photons to two subsequent measurement devices.

      The two detectors in the top device measure the $1^{st}$ qubit (or Particle 1) while the two detectors in the bottom device measure the $2^{nd}$ qubit (or Particle 2). For each device $j$, the controllable parameters $\theta_j$ and $\phi_j$ can be adjusted via the beamsplitters and phase plates exactly as defined in Section \ref{Single Qubit Measurements}. 
     
     It should be noted that the optical path-lengths for each arm of both the devices are exactly equal to each other, i.e. the photons always travel an equal distance to each beamsplitter, no matter which way it is emitted by the source $\bm{E}$. This ensures that no phase difference is introduced due to the path length in each interferometer. Any phase factors being introduced are represented by phase plates.

     It may not be evident that the source pictured Figure \ref{fig:2q} is equivalent to the entangled state $\ket{\psi}$.   The rest of this section will show how the correct QM probabilities can be recovered from a path-integral analysis, proving this equivalence.

   \subsection{Computing Probabilities with the Path Integral} \label{rules for PI}
   
   The rules for combining path integrals (see, \textit{e.g.}, \cite{sinha1991}) are very similar to how one computes classical field amplitudes.  When multiple classical fields arrive at the same point, one adds the complex amplitudes and squares the total to find the classical intensity.  Similarly, when different histories end at the same outcomes, one adds the complex amplitude corresponding to each history and squares the sum to find the probability of those outcomes.  (Recall from the Introduction that a ``history'' corresponds to a single path through configuration space, which in this two-photon case is equivalent to the spacetime paths of the two photons, taken together as a pair.)
   
   Feynman discovered that the amplitude for a history is proportional to $exp(iS/\hbar)$ where $S$ is the classical action \cite{Feynman1948}.  If that history is comprised of multiple paths, as in this two-photon case, the total action is $S_1+S_2$, the sum of the action of the two paths.  Therefore, the total amplitude of a history is $exp(iS_1/\hbar)exp(iS_2/\hbar)$, indicating that one multiplies the amplitude factors from each path in the same history.  This is not to be confused with the rule in the previous paragraph, concerning the case where multiple different histories end at the same outcome; in that case one adds the amplitudes instead of multiplying.  These rules will now be precisely defined in the present context.
   
   To calculate the amplitude for any history in the particular experimental geometry shown in Figure 2, one undertakes the following procedure:
   
  \textbf{1. Source Amplitude}:  Recall that the photons will be emitted on either the two solid lines, or the two dashed lines.  The amplitude of any history beginning on the solid lines is $A$, and the amplitude of any history beginning on the dashed lines is $B$.  Note that this factor is not squared, because the two photons are not independent.  (If these photons were measured by a screen, one would square these amplitudes to determine the joint path probability $A^2$ or $B^2$.)

 \textbf{2. Beamsplitter Factor}: At a beamsplitter, $R$ and $T$ represent the single-photon reflection and transmission probability with the relationship $R + T = 1$.  When a path is reflected at a beamsplitter, the amplitude picks up factor of $i\sqrt{R}$, in precise analogy with the classical field reflection discussed in Section 2.  Similarly, for a transmission we simply pick up the factor of $\sqrt{T}$.  (Note that in our case each history includes paths through both measurement devices, so there are two beamsplitters which will enter into each amplitude calculation.)
   
   \textbf{3. Phase Plate Factor}: Whenever a path in a history passes through a phase plate of angle $\beta$, the amplitude picks up a factor of $e^{i\beta}$.  (This occurs at each circle in Figure \ref{fig:2q}.)
   
  Once the amplitudes $\mathcal{E}$ for each history have been determined, we proceed as described above.  One can bin them into groups having the same final outcomes $k$.  With histories in each outcome-bin labelled by some index $n$, every history has an amplitude of the form $\mathcal{E}^n_k$.  We would like to compute the probability of any particular outcome $k$.  To find this probability, one sums the amplitudes $\mathcal{E}^n_k$ which lead to outcome $k$ and then squares the total;
  \begin{equation}
      \label{eq:masterrule}
      Prob(k) = \left|\sum_n \mathcal{E}^n_k \right|^2 .
  \end{equation}
  In general, a normalization factor will be required, but in the specific case analyzed here, the probabilities always sum to one.

   \subsection{Path Integral Calculations}\label{Path Integral Calculations}
   
   We now will show that the path integral approach described above can be used to calculate all the same probabilities as quantum theory for these arbitrary two-qubit states.  Notice that we will not need to use any non-classical paths or histories; only eight histories need to be considered to get the proper answers.
   
   The reason that there are eight histories is as follows:  since the two photons are emitted in opposite directions (initially travelling along the solid lined paths or the dashed lined paths), there are only two initial possibilities, with amplitudes $A$ and $B$ for the solid and the dashed lines respectively.  For each of these two cases, the photons must either transmit or reflect off a beamsplitter, ending at one detector in each device.  Since there are 4 different outcomes, each one reachable from either of the two initial possibilities, there are eight total histories.
   
   Let's consider in detail the case where both photons reach the $\bm{+}$ detectors.  To find this probability, there are two histories that need to be considered, with photons beginning on both the dashed lines or on both solid lines in Figure \ref{fig:2q}.  
    
    Let's consider the solid-line history first. Particle 1 reaches the $\bm{+}$ detector by getting reflected at the top beamsplitter, while Particle 2 reaches the $\bm{+}$ detector by getting transmitted through the bottom beamsplitter. The amplitude of this history includes $exp(i\phi_1)$ from the top phase plate and $i\sqrt{R_1}$ from the top beamsplitter, and also $i$ from the lower phase plate and $\sqrt{T_2}$ from the bottom beamsplitter. The rules from Section 3.3 therefore yield an amplitude  
    \begin{equation}
        \mathcal{E}^A_{++}={A}{\left(e^{i\phi_1}\:i\:cos\:\frac{\theta_1}{2}\right)\left(i\:sin\:\frac{\theta_2}{2}\right)} .\label{E_A++}
    \end{equation}
   For the other history with the same outcome, tracing the dashed lines to the top and bottom $\bm{+}$ detectors, Particle 1 reaches the $\bm{+}$ detector by getting transmitted through the top beamsplitter while Particle 2 must now be reflected at the bottom beamsplitter.  The top phase plate and top beamsplitter yield a factor $i\sqrt{T_1}$, while Particle 2 passes through two phase plates, thus picking up a factor $exp(i\delta) exp(i\phi_2)$ and then another $i\sqrt{R_2}$ from the reflection. Together these factors yield an amplitude
    \begin{equation}
        \mathcal{E}^B_{++}={B}{\left(i\:sin\frac{\theta_1}{2}\right)\left(e^{i\delta}\:e^{i\phi_2}\:i\:cos\frac{\theta_2}{2}\right)} .\label{E_B++}
    \end{equation}
    Since histories $\mathcal{E}^A_{++}$ and $\mathcal{E}^B_{++}$ are leading to the same outcome, we add them to get the total amplitude and use Eqn. (\ref{eq:masterrule}) to get the probability of the photons arriving at both of the $\bm{+}$ detectors:
    \begin{equation} 
      P(++) = \left|-{A} e^{i\phi_1}\:cos\frac{\theta_1}{2}\:sin\frac{\theta_2}{2} - {B} \:e^{i\delta}\:e^{i\phi_2}\:sin\frac{\theta_1}{2}\:cos\frac{\theta_2}{2}\right|^2 . 
    \end{equation}

    Comparing this probability to  Eqn. (\ref{eqn:qm++}), we see an identical result.  It is evident that a measurement at both $\bm{+}$ detectors corresponds to two qubits measured on the Bloch Sphere at arbitrary angles $\theta_1, \phi_1, \theta_2, \phi_2$ respectively.
    
    This approach also yields the correct probabilities for the other three outcomes.  For the case where Particle 1 is detected at $\bm{+}$ and Particle 2 is detected at $\bm{-}$, the probability is calculated from amplitude $\mathcal{E}^A_{+-}$ (reflecting at both beamsplitters) and $\mathcal{E}^B_{+-}$ (transmitting through both beamsplitters).  This yields
        \begin{equation}
            P(+-)=  \left|{A} i^3\:cos\frac{\theta_1}{2}\:cos\frac{\theta_2}{2}\:e^{i\phi_1} + {B} i\:e^{i\delta}\:sin\frac{\theta_1}{2}\:sin\frac{\theta_2}{2}\:e^{i\phi_2}\right|^2  
        \end{equation} 
    which is equal to the quantum mechanical probability given by Eqn. (\ref{eqn:qm+-}).  
    
     Similarly, combining $\mathcal{E}^B_{-+}$ (reflecting at both the beamsplitters) and $\mathcal{E}^A_{-+}$ (transmitting through both beamsplitters) results in
        \begin{equation}
            P(-+)=  \left|{A} i\:sin\frac{\theta_1}{2}\:sin\frac{\theta_2}{2}\:e^{i\phi_1} + {B} i^3\:e^{i\delta}\:cos\frac{\theta_1}{2}\:cos\frac{\theta_2}{2}\:e^{i\phi_2}\right|^2  ,
        \end{equation} 
     equal to the probability given by Eqn. (\ref{eqn:qm-+}).   
     
     Finally, combining $\mathcal{E}^A_{--}$ (transmitting through the top beamsplitter and reflecting at the bottom beamsplitter), and $\mathcal{E}^B_{--}$ (reflecting at the top beamsplitter and transmitting through the bottom beamsplitter), results in
        \begin{equation}
            P(--)=  \left|{A} i^2\:sin\frac{\theta_1}{2}\:cos\frac{\theta_2}{2}\:e^{i\phi_1} + {B} i^2\:e^{i\delta}\:cos\frac{\theta_1}{2}\:sin\frac{\theta_2}{2}\:e^{i\phi_2}\right|^2  ,
        \end{equation} 
   equal to the probability Eqn. (\ref{eqn:qm--}).        
    
   This concludes the demonstration that any observable correlation of any entangled two-qubit state can be reproduced by a simple sum-over-paths calculation, where each path can be said to have a spacetime representation.  No higher-dimensional configuration space has been required.

\section{A 3-Qubit Example: GHZ}\label{The 3 Qubit case: GHZ}

While Section 3 discussed a generic 2 qubit case, a similar representation is also possible for an entangled 3 qubit case. Instead of a generic 3-qubit state, here we discuss the special case of the GHZ state \cite{greenberger1990}.

\subsection{Quantum Mechanical probabilities}\label{Quantum Mechanical probabilities for ghz}
A 3-qubit state lives in a larger Hilbert space, the tensor product of three single-qubit spaces.  Using the same notation as before, $\ket{\large{\bm{e_+}}}_3$ and $\ket{\large{\bm{e_-}}}_3$ indicate the orthogonal basis chosen by angles $(\theta_3,\phi_3)$ on the Bloch sphere.
For example, a measurement of all three qubits aligned with each chosen basis corresponds to the state
 \begin{eqnarray}
           \ket{\bm{+++}} = \ket{\large{\bm{e_+}}}_1\otimes \ket{\large{\bm{e_+}}}_2\otimes \ket{\large{\bm{e_+}}}_3 = \begin{pmatrix} cos \frac{\theta_1}{2}\:cos\frac{\theta_2}{2}\:cos\frac{\theta_3}{2}\\cos \frac{\theta_1}{2}\:sin\frac{\theta_2}{2}\:cos\frac{\theta_3}{2}\:e^{i\phi_2}\\sin \frac{\theta_1}{2}\:cos\frac{\theta_2}{2}\:cos\frac{\theta_3}{2}\:e^{i\phi_1}\\sin \frac{\theta_1}{2}\:sin\frac{\theta_2}{2}\:cos\frac{\theta_3}{2}\:e^{i(\phi_1 + \phi_2)}\\cos \frac{\theta_1}{2}\:cos\frac{\theta_2}{2}\:sin\frac{\theta_3}{2}\:e^{i\phi_3}\\cos \frac{\theta_1}{2}\:sin\frac{\theta_2}{2}\:sin\frac{\theta_3}{2}\:e^{i(\phi_2 + \phi_3)}\\sin \frac{\theta_1}{2}\:cos\frac{\theta_2}{2}\:sin\frac{\theta_3}{2}\:e^{i(\phi_1 + \phi_3)}\\sin \frac{\theta_1}{2}\:sin\frac{\theta_2}{2}\:sin\frac{\theta_3}{2}\:e^{i(\phi_1 + \phi_2 + \phi_3)} \end{pmatrix}.
 \end{eqnarray}
 The other states ($\ket{\bm{++-}},\ket{\bm{+-+}}$, etc.) can be calculated by taking the other tensor products in a similar manner. 
  The GHZ state is expressed as $\psi=A\ket{000} + \bm{B}\ket{111}$  $=\begin{pmatrix} \bm{A}&0&0&0&0&0&0&\bm{B} \end{pmatrix}^T $. 
 We will focus on the simple case \textbf{A=B=$1/\sqrt{2}$}, although other possibilities are easily addressed using the strategy of unequal emission probabilities from the previous section. The quantum mechanical probabilities for the 8 possible outcome combinations can be calculated using the Born rule:

\begin{equation} \label{qmghz}
\begin{aligned}
          \left|\braket{\psi|{+++}}\right|^2 &=&\!\!\!\frac{1}{2} \left|cos \frac{\theta_1}{2}\:cos\frac{\theta_2}{2}\:cos\frac{\theta_3}{2} + sin \frac{\theta_1}{2}\:sin\frac{\theta_2}{2}\:sin\frac{\theta_3}{2}\:e^{i(\phi_1 + \phi_2 + \phi_3)}\right|^2 \\
          \left|\braket{\psi|{++-}}\right|^2 &=&\!\!\!\frac{1}{2} \left|cos \frac{\theta_1}{2}\:cos\frac{\theta_2}{2}\:sin\frac{\theta_3}{2}\:e^{-i\phi_3} - sin \frac{\theta_1}{2}\:sin\frac{\theta_2}{2}\:cos\frac{\theta_3}{2}\:e^{i(\phi_1 + \phi_2)}\right|^2 \\
          \left|\braket{\psi|{+-+}}\right|^2 &=&\!\!\!\frac{1}{2} \left|cos \frac{\theta_1}{2}\:sin\frac{\theta_2}{2}\:cos\frac{\theta_3}{2}\:e^{-i\phi_2} - sin \frac{\theta_1}{2}\:cos\frac{\theta_2}{2}\:sin\frac{\theta_3}{2}\:e^{i(\phi_1 + \phi_3)}\right|^2 \\
          \left|\braket{\psi|{+--}}\right|^2 &=&\!\!\!\frac{1}{2} \left|cos \frac{\theta_1}{2}\:sin\frac{\theta_2}{2}\:sin\frac{\theta_3}{2}\:e^{-i(\phi_2 + \phi_3)} + sin \frac{\theta_1}{2}\:cos\frac{\theta_2}{2}\:cos\frac{\theta_3}{2}\:e^{i\phi_1}\right|^2 \\
          \left|\braket{\psi|{-++}}\right|^2 &=&\!\!\!\frac{1}{2} \left|sin \frac{\theta_1}{2}\:cos\frac{\theta_2}{2}\:cos\frac{\theta_3}{2}\:e^{-i\phi_1} - cos \frac{\theta_1}{2}\:sin\frac{\theta_2}{2}\:sin\frac{\theta_3}{2}\:e^{i(\phi_2 + \phi_3)}\right|^2 \\
          \left|\braket{\psi|{-+-}}\right|^2 &=&\!\!\!\frac{1}{2} \left|sin \frac{\theta_1}{2}\:cos\frac{\theta_2}{2}\:sin\frac{\theta_3}{2}\:e^{-i(\phi_1 + \phi_3)} + cos \frac{\theta_1}{2}\:sin\frac{\theta_2}{2}\:cos\frac{\theta_3}{2}\:e^{i\phi_2}\right|^2 \\
          \left|\braket{\psi|{--+}}\right|^2 &=&\!\!\!\frac{1}{2} \left|sin \frac{\theta_1}{2}\:sin\frac{\theta_2}{2}\:cos\frac{\theta_3}{2}\:e^{-i(\phi_1 + \phi_2)} + cos \frac{\theta_1}{2}\:cos\frac{\theta_2}{2}\:sin\frac{\theta_3}{2}\:e^{i\phi_3}\right|^2 \\
          \left|\braket{\psi|{---}}\right|^2 &=&\!\!\!\frac{1}{2} \left|sin \frac{\theta_1}{2}\:sin\frac{\theta_2}{2}\:sin\frac{\theta_3}{2}\:e^{-i(\phi_1 + \phi_2+ \phi_3)} - cos \frac{\theta_1}{2}\:cos\frac{\theta_2}{2}\:sin\frac{\theta_3}{2}\right|^2 .
\end{aligned}
\end{equation}

The goal of this section is to demonstrate that these probabilities can be recovered in the path integral formalism.

\subsection{Path Integral Calculation}\label{Path Integral Calculation for ghz}
For the GHZ case the experimental setup is expanded as shown in Figure 3, with three measurement devices instead of two.\footnote{These measurement devices are slightly different than those used in the previous sections, but this allows for a smaller total number of phase plates, simplifying the system.} For this geometry we use two sources ($E_1$ and $E_2$), each emitting two photons as before.  Since we are interested in the case \textbf{A=B=$1/\sqrt{2}$}, we assume these sources have equal probability for emission along either the dashed or solid lines.  To ensure that we are dealing with only three photons, one at each measurement device, the fourth photon is registered by a central detector, labeled by a "$1$" in the figure.  If this central detector measures \textit{exactly} one photon, this signifies that the other three photons are now effectively in the GHZ state, and the measurement can proceed.  In other words, the GHZ state is obtained only after a post-selection that depends on the central detector. Given a successful post-selection, notice that \textit{all} the photons will either be on the solid lines or on the dashed lines, similar to the two-qubit case.
While it might not be evident that this indeed is the GHZ state immediately, the equal quantum mechanical probabilities and path integral calculations proves it.

       \begin{figure}[ht]
        \centering

        \includegraphics[width=0.8\textwidth,height=10.5cm]{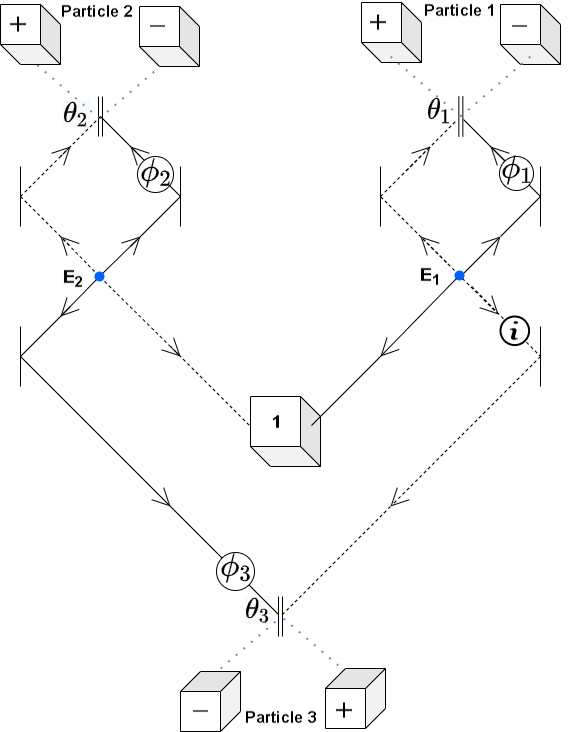}

        \caption{Two sources $\bm{E_1}$ and $\bm{E_2}$ emit two photons each, which travel along the solid or dashed lines. The central detector "$1$" measures \textbf{exactly} one photon, while three measurement devices are used to measure each of the other photons via a locally chosen basis ($\theta,\phi$). } 
        \label{fig:3q}
      \end{figure}
Using the procedure described in Section \ref{rules for PI}, the path integral approach can be applied for any combination of measurement outcomes.  Here we show a derivation of the amplitude for all three photons to be measured at the $\bm{+}$ detectors.  For the case where all photons are on the solid lines, the amplitude is
\begin{equation}
    \mathcal{E}^{solid}_{+++} = \frac{1}{\sqrt{2}}\left(e^{i\phi_1}\:sin\frac{\theta_1}{2}\right)\left((e^{i\phi_2}\:sin\frac{\theta_2}{2})(e^{i\phi_3}\:sin\frac{\theta_3}{2})\right) .   
\end{equation}
This expression is derived as follows.  The solid line path for ${E_1}$ passes through the phase plate $\phi_1$ and gets transmitted through beamsplitter 1, yielding the factor $\sqrt{T_1}=sin(\theta_1/2)$ to reach the $\bm{+}$ detector for particle 1. Two photons emerge from source ${E_2}$.  One photon passes through phase plate $\phi_2$ and gets transmitted through beamsplitter 2, for a factor $\sqrt{T_2}=sin(\theta_2/2)$.  The other passes through phase plate $\phi_3$ and gets transmitted through beamsplitter 3 for a factor of $\sqrt{T_3}=sin(\theta_3/2)$. 

The other possible history is when all photons are on the dashed lines.  In this case, one photon reflects at beamsplitter 1   and passes through a fixed phase plate $i$ yielding a factor $i^2\sqrt{R_1}$.  Another reflects at beamsplitter 3 with $i\sqrt{R_3}$ to reach the $\bm{+}$ detector.  The final photon simply reflects at beamsplitter 2 with a factor of $i\sqrt{R_2}$.  The total amplitude for these three paths is therefore
\begin{equation}
    \mathcal{E}^{dashed}_{+++} = \frac{1}{\sqrt{2}}\left((i\:cos\frac{\theta_1}{2})(i\:cos\frac{\theta_2}{2})\right)\left(i^2\:cos\frac{\theta_3}{2}\right) .  
\end{equation}

As both of these histories lead to the same outcome, they are added and then squared to give the joint probability of all three particles being detected at the $\bm{+}$ detectors.  Specifically,
\begin{eqnarray}
    P(+++) &=&  \left|\mathcal{E}^{dashed}_{+++}+\mathcal{E}^{solid}_{+++}\right|^2 ,\\ 
    &=& \frac{1}{2}\left|\:cos\frac{\theta_1}{2}\:cos\frac{\theta_2}{2}\:cos\frac{\theta_3}{2} + e^{i\phi_1} sin\frac{\theta_1}{2}e^{i\phi_2}\:sin\frac{\theta_2}{2}e^{i\phi_3}\:sin\frac{\theta_3}{2}\right|^2 .   \nonumber      
\end{eqnarray}
By rearranging the equation, this is exactly comparable to the first equation in Eqn. (\ref{qmghz}). 
Similarly, using the amplitudes listed in Appendix B, the probabilities of the other outcomes can also be calculated through path integrals, which turn out to be exactly equal to the corresponding quantum mechanical probabilities calculated in Eqn. (\ref{qmghz}).  Therefore all of the correlations evident from measurements of the GHZ state can be calculated from an analysis of spacetime-based histories.

\section{Discussion} \label{Discussion and Conclusion}

\subsection{Summary}
We begin our discussion with a summary of the main results.  The measurement of a generic single qubit has been developed in Section \ref{Single Qubit Measurements}. The measurement device described in Section \ref{Measurement Device} consists of two adjustable parameters: the phase plate of phase delay angle $\phi$ and the beamsplitter corresponding to an angle $\theta$. These controllable factors are equivalent to choosing an orthogonal basis for the qubit aligned in any direction $(\theta,\phi)$ on the Bloch Sphere. In the single photon limit this device yields the same probabilities as standard quantum mechanics.

Using N copies of this measurement device we can measure N-qubit systems. For a generic 2 qubit case shown in Section \ref{The 2 Qubit case} we would then require two such devices, the geometry of which is discussed in detail in Section \ref{Two-Photon Experimental Geometry}. A key part of the setup was a conceptually simple source that emits two photons in exactly opposite directions, but with different probabilities of emitting along the solid or dashed paths as seen in Figure \ref{fig:2q}. The quantum mechanical probabilities for each possible outcome calculated in Section \ref{Quantum Mechanical probabilities} were shown to be exactly equal to the probabilities calculated via the path integral approach described in Section \ref{Path Integral Calculations}. It must be noted that these calculations are carried out by following the path integral procedure described in Section \ref{rules for PI}, and correctly recovered all observable correlations for any entangled state.  

The special 3 qubit case of the GHZ state followed the same procedure. Two sources were required to create this state, where each source emits two photons in exactly opposite directions, as shown in Figure \ref{fig:3q}. Unlike the generic two-qubit case, these sources emit with the same probability in all directions. A version of the measurement device was used for each of the three photons, while the fourth photon was registered by a special central detector. (If this fourth detector does not see a single photon, the GHZ state is not created.)  The quantum probabilities calculated in Section \ref{Quantum Mechanical probabilities for ghz} corresponded exactly with the path integral calculations carried out in Section \ref{Path Integral Calculation for ghz}.

\subsection{Analysis}

Formally, a path integral calculation requires a sum over all possible histories which lead to the same endpoint, but that is evidently not the case in the above examples.  As we saw, only a few classical paths were required.\footnote{Here we are using the word ''classical'' to indicate that only straight-line paths are required.  If a photon is thought of as an electromagnetic wave, it is not literally classical for that wave to entirely pass through a beamsplitter (or to entirely reflect), but the path integral formalism does not evidently allow for including both beamsplitter outputs into a single history if only one detector fires.}  Because of this simplification, the difficulty of the calculation in the path integral framework is not essentially more difficult than the traditional quantum calculation.  Indeed, since no tensor products or inner products need to be calculated, the number of formal calculations seems somewhat reduced in the path-integral approach, especially for the 3-qubit case.  If the number of qubits were to increase beyond three, we expect this advantage to be even more pronounced.  
Of course, the difficulty of the calculation grows regardless; even in the path integral framework one would need to increase the number of measurement devices and create a setup with the appropriate number of photon sources.

More important than the number of calculational steps, however, is the fact that the histories used by the path integral framework reside in spacetime rather than in an abstract Hilbert space.  Even for entangled states, each possible history can be mapped to physical paths of particles which travel from the sources to the detectors.  (These paths end at only the detectors which actually register a particle, never at the detectors which do not.)  There is no comparable spacetime-based picture when analyzing entangled states in the conventional QM formalism.

Some readers may be resistant to this conclusion, perhaps because of an impression that Bell's Theorem prevents a spacetime-based account of entanglement experiments.  Even setting aside the fact that this view of Bell's Theorem is not true in general \cite{Wharton2020}, it is simple to see why the path integral framework does not fall within Bell's analysis, as Bell assigned probabilities $Prob(\lambda)$ to hidden/unknown parameters $\lambda$.  The path integral framework does not assign \textit{probabilities} to each spacetime-based history, but rather complex \text{amplitudes}.  These amplitudes are then combined in a way that is incompatible with a probabilistic assignment.  For example, if one history had an amplitude of $iA$, and another history had an amplitude of $-iA$, the square of the sum of these amplitudes is zero, corresponding to a zero probability.  This would be impossible for two hidden possibilities, each with a non-zero probability, and it results from the curious way in which amplitudes are combined; see \cite{Wharton2016} for more discussion of this issue.

But the above resolution to one concern might raise another.  If the amplitudes for each spacetime-based history are combined in a way that is inconsistent with hidden spacetime-based parameters $\lambda$, then is it really fair to say that this is a spacetime-based calculation?  While these histories do not directly link distant points in space, as conventional quantum superpositions, it is indeed true that {different} histories are being combined into the same probability calculation in some non-classical manner.  (This must be the case, or else it would not be possible to violate Bell Inequalities in the first place.)

Still, it is certainly true that alternate perspectives of the same phenomena have historically been found to generate useful advances and generalizations.  In this case, if one is concerned about the disconnect between entangled states and conventional spacetime, the path integral is an obvious viewpoint which might partially resolve this tension.  In particular it does not require any direct connection between spacelike-separated regions of spacetime, and is therefore at least somewhat more compatible with classical spacetime than are instantaneous entangled states.  Additional applications of this alternate perspective on entanglement will now be discussed.  

\subsection{Potential Applications}

The results from Sections 3 and 4 provide a standard recipe for applying the alternate path-integral viewpoint to many well-known results in quantum theory and quantum foundations.  Take, for example, the no-signalling theorem \cite{ghirardi1980}.  This theorem is formally developed in a conventional QM framework, but it must also somehow be implied by the path-integral formalism.  On its face, such a theorem seems more surprising in the path integral framework, where adjusting measurement settings ($\theta, \phi$) on one detector changes the amplitude for entire histories that include distant measurement outcomes.  But, since the joint probabilities are identical to that of conventional QM, even the path integral must respect this no-signalling condition; the marginal probabilities for distant outcomes must always cancel when local devices are adjusted.  Exploring why this is necessarily the case seems like it might provide further insight into how signal locality is necessarily enforced, even in an ``all at once'' framework like the path integral.

A related application would be the question of ``fine-tuning'' \cite{Wood2015}, and whether there is a natural explanation for the perfect cancellation of the marginal probabilities.  Analysis of this issue in a path-integral framework might indicate whether this cancellation might be explained via symmetries \cite{Almada2016}, global consistency \cite{Adlam2018}, or some other factor that becomes evident from this alternate perspective.

Another evident line of future research would follow from a basic extension of these results: time-reversing the entire set-up in Figure 2, where the detectors become single-photon sources and the central two-photon emitter $\bm{E}$ becomes instead a possible two-photon absorber.  Treating other two-photon processes at $\bm{E}$ (scattering, etc.) as alternate possible outcomes, one could likely build the analog of a two-photon measurement device.  With this in hand, it would be possible to apply the path integral framework to other known results in quantum theory; Hanbury-Brown-Twiss experiments, entanglement swapping, the geometry used in the Pusey-Barrett-Rudolph theorem \cite{Pusey2012}, etc.  In every case, one expects the same results as QM, but for evidently different \textit{reasons}, and these reasons may very well shed light on outstanding questions in quantum foundations.  

Note that a further advantage to such a path integral framework is that the accounts of these experiments would be far more evidently time-symmetric than the conventional QM viewpoint \cite{Wharton2011}.  In conventional QM, time-reversing a two-particle Bell experiment transforms it into a geometry where the particles are not even considered to be entangled.  But a time-reversed history in the path-integral viewpoint does not change its amplitude whatsoever.  There are still differences between these two experiments, even in the path-integral viewpoint, but those differences come in via the experimenter (what the experimenter can be said to control is different in these two cases) rather than the mathematics that purportedly describes the quantum system. 

Finally, it should again be noted that the path integral is the obvious starting point for any reformulation of quantum theory built upon the ``all at once'' analysis familiar from general relativity (where spacetime systems are solved as coherent entities rather than computed sequentially).  If one takes entanglement phenomena to be one of the most difficult-to-reformulate aspects of quantum theory, then a path-integral view of entanglement is a necessary starting point for any ``all at once'' account of QM.  Any resulting model of these phenomena would almost certainly fall into the ``Future Input Dependent'' class of models described in a recent review \cite{Wharton2020}.

Even apart from the above potential applications, these results clearly demonstrate not merely that a path integral approach \textit{can} recover the usual joint probabilities for measurements of entangled states, but precisely \textit{how} to calculate these probabilities for any pure two-qubit state.  The analysis does not require any computations in Hilbert space, and can be performed entirely using histories built out of spacetime-based paths, with no direct connections between spacelike-separated events.  Using the above three-qubit GHZ state as a guide, we expect that further generalizations to a path integral analysis of any N-qubit entangled state should be conceptually straightforward, although the details of how to accomplish this remains an open problem.

\section*{Appendix A: The Schmidt Basis}\label{AppA}

For some applications of the two-qubit path integral, one may want to start with a completely general two-qubit pure state
\begin{equation}
\label{eq:generic}
\ket{\psi}=a\ket{\bar{0}\bar{0}}+b\ket{\bar{0}\bar{1}}+c\ket{\bar{1}\bar{0}}+d\ket{\bar{1}\bar{1}},
\end{equation}
where $a,b,c$ and $d$ are complex and $|a|^2 + |b|^2 + |c|^2 + |d|^2 = 1$.  Here the overbars on $\ket{\bar{0}}$ and $\ket{\bar{1}}$ are to distinguish these states from the Schmidt basis $\ket{{0}}$ and $\ket{{1}}$ used in Section 3.  

Each of the two qubits in $\ket{\psi}$ can then be measured in some arbitrary direction on the Bloch sphere, corresponding to the angles $(\bar{\theta}_1,\bar{\phi}_1)$ for the first qubit and $(\bar{\theta}_2,\bar{\phi}_2)$ for the second.  These angles are defined relative to the corresponding basis on the right side of (\ref{eq:generic}), and will generally be different than the angles $(\theta_1,\phi_1)$ and $(\theta_2,\phi_2)$ used in Section 3.  (These latter angles are defined relative to the Schmidt basis states $\ket{{0}}$ and $\ket{{1}}$.)

However, it is a relatively straightforward manner to relate these various coordinate systems, if one can solve for the direction $(\alpha,\beta)$ on the given Bloch sphere which lines up with the z-axis of the Schmidt basis Bloch sphere.  For the first qubit, this relationship is defined by
\begin{eqnarray}
\ket{0}_1 &=& \cos \frac{\alpha_1}{2} \ket{\bar{0}}_1 + \sin \frac{\alpha_1}{2} e^{i\beta_1} \ket{\bar{1}}_1\\
\ket{1}_1 &=& \sin \frac{\alpha_1}{2} e^{-i\beta_1} \ket{\bar{0}}_1 - \cos \frac{\alpha_1}{2}  \ket{\bar{1}}_1.
\end{eqnarray}
And a similar (but reversed) expression for the second qubit is
\begin{eqnarray}
\ket{0}_2 &=& \sin \frac{\alpha_2}{2} e^{-i\beta_2} \ket{\bar{0}}_2 - \cos \frac{\alpha_2}{2}  \ket{\bar{1}}_2\\
\ket{1}_2 &=& \cos \frac{\alpha_2}{2} \ket{\bar{0}}_2 + \sin \frac{\alpha_2}{2} e^{i\beta_2} \ket{\bar{1}}_2.
\end{eqnarray}

Recall, that if written in the Schmidt basis, the state $\ket{\psi}$ takes the much simpler form given by Eqn. (\ref{eq:psi2}).  If one knows the angles $(\alpha,\beta)$, then one knows precisely how the measurement angles $(\bar{\theta},\bar{\phi})$ in the given basis are related to the measurement angles $(\theta,\phi)$ in the Schmidt basis.  Geometrically, on the Bloch sphere, they are related by a rotation which takes $(\alpha,\beta)$ to the z-axis, but any careful analysis would certainly want to use the precise above equations.

All that remains is to find the four angles $\alpha_1$, $\beta_1$, $\alpha_2$ and $\beta_2$.  We have not been able to find a derivation of these angles in the published literature, but they can essentially be recovered from the results in \cite{wharton2016b}, except for the special case of maximally-entangled states which will be worked out separately.  In terms of the four complex parameters in Eqn. (\ref{eq:generic}), these angles are

\begin{equation}
    \cos \alpha_1 = \frac{|a|^2+|b|^2-|c|^2-|d|^2}{\sqrt{1-4|ad-bc|^2}},
\end{equation}

\begin{equation}
    e^{-i\beta_1}=\frac{ac^*+bd^*}{|ac^*+bd^*|},
\end{equation}

\begin{equation}
    \cos \alpha_2 = \frac{|a|^2+|c|^2-|b|^2-|d|^2}{\sqrt{1-4|ad-bc|^2}},
\end{equation}

\begin{equation}
    e^{-i\beta_2}=\frac{ab^*+cd^*}{|ab^*+cd^*|}.
\end{equation}

Via tedious algebra, all of the above equations in this appendix can be combined to ascertain that both $\braket{\psi|00}$ and $\braket{\psi|11}$ are exactly zero.  (We have also checked this numerically.)  This algebra is much easier if one uses the normalization of $\ket{\psi}$ to prove the relationships

\begin{equation}
    \sin \alpha_1 = \frac{2|ac^*+bd^*|^2}{\sqrt{1-4|ad-bc|^2}},
\end{equation}

\begin{equation}
    \sin \alpha_2 = \frac{2|ab^*+cd^*|^2}{\sqrt{1-4|ad-bc|^2}}.
\end{equation}

These equations also indicate what is going on when one of the $\beta$ terms become undefined -- that is, when either $|ac^*+bd^*|$ or $|ab^*+cd^*|$ is zero.  In these cases, the corresponding $\alpha$ also goes to zero or $\pi$, such that the two bases are aligned for that qubit. (For $\alpha=0$ it is already in the Schmidt basis, and for $\alpha=\pi$ one need merely set $\ket{0}=\ket{\bar{1}}$ and $\ket{1}=\ket{\bar{0}}$.)  The fact that $\beta$ is undefined is therefore just the usual spherical-coordinate ambiguity for $\phi$ on the z-axis; in this case $\beta$ can take any value, such as $\beta=0$.

The only remaining problem is the special case of maximally-entangled states, where ${{1-4|ad-bc|^2}}=0$; evidently in this case the above angles $\alpha$ are undefined.  In that case, it is always possible to leave one of the qubits unchanged, and then find a Schmidt basis using only the other qubit.  These maximally entangled states can always \cite{kus} be written in the form

\begin{eqnarray}
\label{eq:gammachichi}
    \bm{\ket{\psi}}=\frac{1}{\sqrt{2}}\left[e^{i\chi_1}cos\gamma\!\ket{\bar{0}\bar{0}}+e^{i\chi_2}sin\gamma\!\ket{\bar{0}\bar{1}}+e^{-i\chi_2}sin\gamma\!\ket{\bar{1}\bar{0}}-e^{-i\chi_1}cos\gamma\!\ket{\bar{1}\bar{1}}\right].
\end{eqnarray}

For this general maximally-entangled state, if one leaves the first qubit unchanged, $\bar{\ket{0}}_1=\ket{0}_1$, the second qubit can be written as a superposition of a Schmidt basis $\ket{0}_2$ and $\ket{1}_2$
\begin{eqnarray}
    \bar{\ket{0}}_2=f\ket{0}_2+g\ket{1}_2, \\
    \bar{\ket{1}}_2=g^*\ket{0}_2-f^*\ket{1}_2.
\end{eqnarray}

For a Schmidt basis of the form Eqn.(\ref{eq:psi2}), we require $\braket{\psi|00}=0$ and $\braket{\psi|11}=0$.  Therefore
\begin{eqnarray}
    (e^{i\chi_1}cos\gamma)^*f+(e^{i\chi_2}sin\gamma)^*g=0,\\
    (e^{-i\chi_2}sin\gamma)^*g^*-(-e^{-i\chi_1}cos\gamma)^*f^*=0.
\end{eqnarray}
As the coefficients $f$ and $g$ are normalized,we get: 
\begin{eqnarray}
    f&=&sin\gamma,\\
    g&=&-e^{-i(\chi_1-\chi_2)}\:cos\gamma.
\end{eqnarray}

The above results can be combined to indicate precisely how to rotate the coordinate systems of the measurement devices, such that the results from Section 3 can directly apply to any possible pure two qubit state.

\section*{Appendix B: GHZ Details}\label{AppB}

 In this Appendix, we provide the details of the GHZ path integral calculations for the other possible outcomes. (Calculations for the $\bm{+++}$ outcome have been discussed in detail in Section \ref{Path Integral Calculation for ghz}.) Using Eqn.(\ref{eq:masterrule}) one can recover the probabilities in Eqn.(\ref{qmghz}) 

For each outcome there are two histories that lead to it; the amplitudes for these histories are listed below and have been calculated using the procedure outlined by Section \ref{rules for PI} for Figure (\ref{fig:3q}).
In each of these cases the first parentheses corresponds to particle 1, the second corresponds to particle 2 and the third corresponds to particle 3.
\begin{eqnarray}
    \begin{aligned}
      \mathcal{E}^{solid}_{++-} &= \frac{1}{\sqrt{2}}\left(e^{i\phi_1}\:sin\frac{\theta_1}{2}\right)\left(e^{i\phi_2}\:sin\frac{\theta_2}{2}\right)\left(i\:e^{i\phi_3}\:cos\frac{\theta_3}{2}\right), \\
     \mathcal{E}^{dashed}_{++-} &= \frac{1}{\sqrt{2}}\left(i\:cos\frac{\theta_1}{2}\right)\left(i\:cos\frac{\theta_2}{2}\right)\left(i\:sin\frac{\theta_3}{2}\right), \\
     \mathcal{E}^{solid}_{+-+} &= \frac{1}{\sqrt{2}}\left(e^{i\phi_1}\:sin\frac{\theta_1}{2}\right)\left(e^{i\phi_2}\:i\:cos\frac{\theta_2}{2}\right)\left(e^{i\phi_3}\:sin\frac{\theta_3}{2}\right),  \\
    \mathcal{E}^{dashed}_{+-+} &= \frac{1}{\sqrt{2}}\left(i\:cos\frac{\theta_1}{2}\right)\left(sin\frac{\theta_2}{2})\right)\left(i^2\:cos\frac{\theta_3}{2}\right),\\
    \mathcal{E}^{solid}_{+--} &= \frac{1}{\sqrt{2}}\left(e^{i\phi_1}\:sin\frac{\theta_1}{2}\right)\left(e^{i\phi_2}\:i\:cos\frac{\theta_2}{2}\right)\left(e^{i\phi_3}\:i\:cos\frac{\theta_3}{2}\right) ,\\
    \mathcal{E}^{dashed}_{+--} &= \frac{1}{\sqrt{2}}\left(i\:cos\frac{\theta_1}{2}\right)\left(sin\frac{\theta_2}{2}\right)\left(i\:sin\frac{\theta_3}{2}\right), \\
    \mathcal{E}^{solid}_{-++} &= \frac{1}{\sqrt{2}}\left(e^{i\phi_1}\:i\:cos\frac{\theta_1}{2}\right)\left(e^{i\phi_2}\:sin\frac{\theta_2}{2}\right)\left(e^{i\phi_3}\:sin\frac{\theta_3}{2}\right),  \\
    \mathcal{E}^{dashed}_{-++} &= \frac{1}{\sqrt{2}}\left(sin\frac{\theta_1}{2}\right)\left(i\:cos\frac{\theta_2}{2}\right)\left(i^2\:cos\frac{\theta_3}{2}\right), \\   
    \mathcal{E}^{solid}_{-+-} &= \frac{1}{\sqrt{2}}\left(e^{i\phi_1}\:i\:cos\frac{\theta_1}{2}\right)\left(e^{i\phi_2}\:sin\frac{\theta_2}{2}\right)\left(e^{i\phi_3}\:i\:cos\frac{\theta_3}{2}\right) ,\\  
    \mathcal{E}^{dashed}_{-+-} &= \frac{1}{\sqrt{2}}\left(sin\frac{\theta_1}{2}\right)\left(i\:cos\frac{\theta_2}{2}\right)\left(i\:sin\frac{\theta_3}{2}\right) ,\\      
    \mathcal{E}^{solid}_{--+} &= \frac{1}{\sqrt{2}}\left(e^{i\phi_1}\:i\:cos\frac{\theta_1}{2}\right)\left(e^{i\phi_2}\:i\:cos\frac{\theta_2}{2}\right)\left(e^{i\phi_3}\:sin\frac{\theta_3}{2}\right), \\  
    \mathcal{E}^{dashed}_{--+} &= \frac{1}{\sqrt{2}}\left(sin\frac{\theta_1}{2}\right)\left(i\:cos\frac{\theta_2}{2}\right)\left(i\:sin\frac{\theta_3}{2}\right), \\    
    \mathcal{E}^{solid}_{---} &= \frac{1}{\sqrt{2}}\left(e^{i\phi_1}\:i\:cos\frac{\theta_1}{2}\right)\left(e^{i\phi_2}\:i\:cos\frac{\theta_2}{2}\right)\left(e^{i\phi_3}i\:cos\frac{\theta_3}{2}\right),\\  
    \mathcal{E}^{dashed}_{---} &= \frac{1}{\sqrt{2}}\left(sin\frac{\theta_1}{2}\right)\left(sin\frac{\theta_2}{2}\right)\left(i\:sin\frac{\theta_3}{2}\right).
    \end{aligned}
\end{eqnarray}

Plugging these amplitudes into Eqn.(\ref{eq:masterrule}) yields the probabilities given by Eqn.(\ref{qmghz}).

\section*{Acknowledgements}
The authors gratefully thank Emily Adlam and Hilary Hurst for helpful suggestions.

\bibliographystyle{unsrt}
\bibliography{PathIntBib.bib}

\end{document}